# Solar Cycle Variation of Sustained Gamma Ray Emission from the Sun

Nat Gopalswamy[1], Pertti Mäkelä[1,2], Seiji Yashiro[1,2], Sachiko Akiyama[1,2], Hong Xie[1,2], and G. Sindhuja[1,2,3]

[1]NASA Goddard Space Flight Center, Greenbelt, Maryland

[2]The Catholic University of America, Washington, DC

[3]Institute of Experimental and Applied Physics, Kiel University, Kiel, Germany

## ABSTRACT

We investigated the occurrence rate of the sustained gamma ray emission (SGRE) events from the Sun using data obtained by the Large Area Telescope (LAT) on board the Fermi satellite since its launch in 2008. The data cover the whole of solar cycle (SC) 24 and the rising and maximum phases of SC 25. One of the challenges was to estimate the number of SGRE events in SC 25 because of a solar array drive assembly's malfunction starting in March 2018 that resulted in a large reduction in solar coverage (LAT gap). This is likely the reason for the small number (16) of SGRE events observed during the first 61 months of SC 25, whereas 27 events were observed during the weaker SC 24 over the corresponding epoch. Over the first 61 months, the average sunspot number (SSN) increased from 56.9 in SC 24 to 79.0 in SC 25. Other energetic events closely related to SGREs, viz., fast and wide (FW) coronal mass ejections (CMEs) and decameter-hectometric (DH) type II bursts also increased significantly in SC 25 by 29% and 33%, respectively when normalized to SSN. Therefore, the increase in solar activity should result in a higher number of SGREs in SC 25. We estimated the number of SGREs in SC 25 using three methods. (i) If the SGRE number varies commensurate with SSN, we should have 38 (27×1.39) SGRE events in SC 25 based on the 27 events in SC 24. However, FW CMEs and DH type II bursts in SC 25 were overabundant by 29% and 33%, so the number SGRE events should be accordingly higher: 48 (27×1.39×1.29 from FW CMEs) and 50 (27×1.39×1.33). (ii) In SC 24, ≈18% of FW CMEs and 27% of DH type II bursts were associated with SGRE events. If the same association rate prevails in SC 25, we should have 48 and 49 SGRE events in this cycle. (iii) Since SGRE events are invariably associated with > 100 keV hard X-ray (HXR) bursts, we identified DH type II bursts associated with > 100 keV HXR bursts from Fermi's Gamma-ray Burst Monitor (GBM) during LAT data gaps. Based on our finding that SGRE events in SCs 24 and 25 were all associated with HXR bursts of duration $\gtrsim$ 5 min, we found only 27 of the 79 LAT-gap type II bursts had > 100 keV HXR bursts with duration $\gtrsim$ 5 min. These DH type II bursts are likely to indicate SGRE, bringing the total number of SGRE events to 43 (16 observed + 27 inferred). Thus, the three methods provide similar estimates of the number of SGRE events in SC 25. We, therefore, conclude that that SC 25 is stronger than SC 24 based on the estimated number SGRE events. Other energetic phenomena such as halo CMEs, ground level enhancement (GLE) events, and intense geomagnetic storms are also consistent with a stronger SC 25.

## 1. Introduction

The level of solar activity is a key factor that decides the frequency and intensity of solar eruptions (flares and coronal mass ejections (CMEs)) and their space weather consequences. CMEs result in large solar energetic particle (SEP) events via the shock they drive and cause intense geomagnetic storms when CMEs directly impinge upon Earth's magnetic field. The level of solar activity is typically quantified using the sunspot number (SSN). High SSN indicates a high abundance of closed magnetic field regions, which are seats of solar eruptions. Therefore, one expects more energetic events such as ultrafast CMEs and major solar flares during solar maxima. During solar minima, eruptions are weak and occur only from quiescent filament regions. Energetic phenomena with severe space weather consequences show long-term variability in their occurrence rate somewhat similar to SSN: fast and wide (FW) CMEs, halo CMEs, large SEP events, intense geomagnetic storms, interplanetary (IP) type II radio bursts, coronal and IP shocks, and IP CMEs (ICMEs). FW CMEs are defined as those with speed ≥ 900 km $s^{-1}$ and width ≥ 60° (Gopalswamy et al. 2001). Large SEP events are those with proton intensity ≥ 10 pfu in the > 10 MeV GOES energy channel. An intense geomagnetic storm is characterized by the disturbance storm time (Dst) index ≤ -100 nT. Solar activity also modulates the physical conditions in the heliosphere. For example, the total pressure in the heliosphere is low during weak solar cycles affecting the physical properties of CMEs (Gopalswamy et al. 2014; 2015a; 2023a). Two phenomena that show variations over and above that in SSN are ground level enhancement (GLE) in SEPs and intense geomagnetic storms.

Solar Cycles (SCs) 23 and 24 had the most complete observations of CMEs, considered to be key sources of SEPs and geomagnetic storms. Cycle 24 proved to be the weakest solar cycle in the Space Age. The weak activity resulted in mild space weather in SC 24: the number of large SEP events and intense geomagnetic storms decreased by 74% and 55%, respectively relative to SC 23. These decreases are significantly larger than the 39% reduction in SSN and 48% decrease in the number of FW CMEs (Gopalswamy et al. 2022). The number of GLE events showed extreme reduction: 16 in SC 23 vs. just 2 in SC 24 (i.e., by 87%). Thus, the highest energy particle events seem to be a very sensitive indicator of the cycle strength.

A phenomenon closely related to GLE events is the sustained gamma-ray emission (SGRE) from the Sun. SGRE was first recognized by Forrest et al. (1985) as emission lasting beyond the impulsive phase of the associated solar flare. Only a handful of SGRE events were recorded from the time of their discovery until the advent of the Large Area Telescope (LAT, Atwood et al. 2009) on board the Fermi satellite (Thompson and Wilson-Hodge, 2022). A key characteristic of SGRE events is their association with the most energetic CME population, similar to that causing GLEs (Gopalswamy et al. 2018). The connection is understandable because the underlying particles are accelerated by the same CME-driven shock. Thus, GLEs and SGRE are indicators of the highest energy particles accelerated by CME-driven shocks. In order to detect GLEs, the observer needs to be well-connected to the eruption region, so the high-energy particles can readily travel to the detector. This is the reason GLEs are generally detected from the western hemispheric eruptions, that too when the source is located close to the ecliptic (Gopalswamy et al. 2013; Gopalswamy and Mäkelä 2014; Gopalswamy et al. 2021). The lack of magnetic connectivity, therefore, is one of the main reasons for the rarity of GLE events. Fortunately, as electromagnetic radiation, SGRE is not affected by magnetic connectivity. This has been demonstrated by the occurrence of SGRE events from all across the solar disk, including from poorly connected east-limb regions (Share et al. 2018; Gopalswamy et al. 2019a; Ajello et al. 2021), and CMEs heading at large angles from the ecliptic. Therefore, SGRE events are better indicators of energetic eruptions that accelerate >300 MeV protons.

Since SGREs are closely related to energetic CMEs from active regions, we expect them to occur in greater numbers in SC 25 because this cycle is stronger than SC 24 (e.g., Nandy 2021). However, a preliminary investigation showed that the number of SGRE events declined in SC 25 by ≈ 40%, while the SSN increased by the same amount over the rise to maximum phases in each cycle (Gopalswamy et al. 2025a). The SGRE trend is also opposite to that of related phenomena such as decameter-hectometric (DH) type II radio bursts and FW CMEs. Gopalswamy et al. (2025a) concluded that ≈ 3 times more SGRE events are expected than the observed 16 events in SC 25. The reduction in the observed number of SGRE events seems to be due to a mechanical problem with the solar array drive assembly (SADA) that started on March 16, 2018. Because of this problem, SADA was not able to rotate the array for continuous illumination by the Sun. Following the anomaly, the Sun is kept toward the edge of the LAT field of view (FOV) so that the array remains illuminated during orbit day. Such an arrangement reduced the Sun exposure resulting in large Fermi/LAT data gaps (DGs) extending over a week at a time. The purpose of this article is to estimate the number of SGRE events that might have occurred during these gaps and hence check the strength of SC 25 relative to SC 24.

In order to estimate the number of SGRE events that might have occurred during Fermi/LAT gaps, we shall make use of the close connection among hard X-ray (HXR) bursts, SEP events, and FW CMEs (Garcia 1994; Kiplinger 1995; Ling and Kahler 2020; Kahler and Ling 2020). We extend this connection to DH type II bursts. Share et al. (2018) reported that all SGRE events in SC 24 they studied were associated with an impulsive phase HXR burst at energies > 100 keV and suggested that such HXR emission is a necessary condition for SGRE. Flares with <100 keV HXR bursts were not associated with SGRE, had lower soft X-ray power, slower CMEs, and weaker metric type II burst association; the HXR bursts also had steeper spectra. They examined a control sample of ≈ 95 events that met at least one of the following criteria: (i) presence of a fast CME (speed ≥ 800 km s$^{-1}$), (ii) association with SEP events with peak flux ≥ 1 pfu in the >10 MeV integral channel, and (iii) involvement of HXR burst at energies > 100 keV. A subset of 19 events satisfied criteria (i) and (iii), out of which 14 (or 74%) were associated with SGRE. They could not rule out the presence of an SGRE in the remaining 5 events because they were all limb events (in which the observed gamma-ray flux is greatly reduced, see Gopalswamy et al. 2021) and the Fermi/LAT duty cycle was unfavorable. They suggested that > 100 keV HXR emission may indicate flare-site acceleration of sub-MeV protons that serve as seed particles for further acceleration by the accompanying CME-driven shock. This close association between impulsive > 100 keV HXR bursts and SGRE is significant for our investigation. We shall make use of this HXR criterion (Share et al. 2018) to estimate the number of SGRE events that might have occurred during Fermi/LAT data gaps.

## 2. Observational Results

SGRE events from SC 24 Fermi/LAT observations have been compiled and reported in a number of catalogs (Allafort et al. 2018; Share et al. 2018; Gopalswamy et al. 2019a; Ajello et al. 2021). For SC 25, we started with the SGRE candidates (https://hesperia.gsfc.nasa.gov/fermi_solar/) compiled by the Fermi team. We confirmed these events using light curves and compiled related energetic phenomena such as soft/hard X-ray flares, CMEs, SEP events, and type II radio bursts. We also use Fermi's Gamma-ray Burst Monitor (GBM, Meegan et al. 2009) data from http://hesperia.gsfc.nasa.gov/fermi/gbm/qlook/orbit_plots/ in examining the hard X-ray flare activity during the study period. These data are particularly important for identifying flares during Fermi/LAT data gaps. Finally, we also use > 100 keV HXR information from Konus-Wind (https://www.ioffe.ru/LEA/kw/index.html, Aptekar et al. 1995) to determine the onset times of some partially observed GBM bursts. In doing so, we make use of the information available at the Coordinated Data Analysis Workshop (CDAW) data center (https://cdaw.gsfc.nasa.gov): the CME catalog consisting of all CMEs manually identified (https://cdaw.gsfc.nasa.gov/CME_list, Yashiro et al. 2004; Gopalswamy et al. 2009; Gopalswamy et al. 2024). We also use the halo CME catalog (https://cdaw.gsfc.nasa.gov/CME_list/halo/halo.html, Gopalswamy et al. 2010), which provides additional information such as deprojected speed and CME source location on the Sun. These two catalogs are compiled primarily from images obtained by the Large Angle and Spectrometric Coronagraph (LASCO, Brueckner et al. 1995) on board the Solar and Heliospheric Observatory (SOHO) mission. Type II radio bursts in the DH wavelength are indicative of CME-driven shocks in the corona and IP medium. These bursts as observed by the radio and plasma wave (WAVES) experiment on board the Wind (Bougeret et al. 1995) and the Solar Terrestrial Relations Observatory (STEREO, Bougeret et al. 2008) have been compiled and listed (https://cdaw.gsfc.nasa.gov/CME_list/radio/waves_type2.html, Gopalswamy et al. 2019b). This catalog features radio dynamic spectra showing type II bursts and the associated phenomena such as flares and CMEs. Large SEP events (proton intensity ≥10 pfu in the >10 MeV energy channel) obtained from GOES proton data and the associated eruptive events are available at https://cdaw.gsfc.nasa.gov/CME_list/sepe/.

We use the sunspot number (SSN) V2.0 (World Data Center - Sunspot Index and Long-term Solar Observations (WDC-SILSO), Royal Observatory of Belgium, Brussels, DOI: https://doi.org/10.24414/qnza-ac80) to derive an average SSN over the first 61 months in SCs 24 and 25.

**Table 1.** Energetic events associated with SGRE during the first 61 months of Cycles 24 and 25

| Property | SC 24[b] | SC 25[b] | SC25/SC24 |
|---|---|---|---|
| Averaged SSN | 56.9 | 79.0 | 1.39 |
| No. Halo CMEs | 192 (3.37)[a] | 247 (3.12) | 1.29 (0.93) |
| No. FW CMEs | 149 (2.62) | 268 (3.39) | 1.80 (1.29) |
| No. DH Type II bursts | 99 (1.74) | 183 (2.32) | 1.85 (1.33) |
| No. Intense magnetic storms | 12 (0.21) | 18 (0.23) | 1.5 (1.06) |
| No. ≥M1.0 flares | 389 | 1525 | 3.92 (2.82) |
| No. ≥10 pfu SEP events | 30 (0.53) | 35 (0.44) | 1.17 (0.83) |
| No. GLE events | 1 (0.02) | 4 (0.05) | 4.0 (2.50) |
| No. DH Type II bursts | 99 (1.74) | 183 (2.32) | 1.85 (1.33) |
| No. SGRE events | 27 (0.47) | 16 (0.20) | 0.59 (0.43) |
| No. SGRE/No. FW CMEs | 0.18 | 0.06 | 0.33 |
| No. SGRE/No. DH type IIs | 0.27 | 0.09 | 0.33 |
| No. SGRE/No. SEP events | 0.90 | 0.46 | 0.51 |

[a]The numbers in parentheses are normalized to the corresponding SSN
[b]First 61 months of SC 24 (2008 December 1 to 2013 December 31) and SC 25 (2019 December 1 to 31 December 2024) are compared.

### 2.1 Energetic Events in SCs 24 and 25

Table 1 compiles the number of energetic events (halo CMEs, FW CMEs, DH type II bursts, intense geomagnetic storms, major soft X-ray flares (flare class ≥ M1.0), large SEP events, and GLE events). These numbers are compared between SCs 24 and 25 over the first 61 months in each cycle (2008 December 1 to 2013 December 31 in SC 24; 2019 December 1 to 31 December 2024 in SC 25). The monthly mean SSN is averaged over the first 61 months in each cycle and used as reference. Also listed are the number of SGRE events over the same epochs in the two cycles. We have compared the numbers directly and by normalizing them to SSN. Normalization accounts for the dependence of the number of events on solar activity. Table1 shows that the average SSN in SC 25 is 79.0, increasing from 56.9 in SC 24 pointing to a stronger SC 25 by 39%.

When CMEs appear to surround the occulting disk of the observing coronagraph in sky-plane projection, they are called halo CMEs (Howard et al. 1982; 1985; Gopalswamy et al. 2010). Halo CMEs have been found to be a sensitive indicator of solar cycle strength in that weak cycles have a higher halo CME abundance (Gopalswamy et al. 2015b; 2023a). There are clearly more halo CMEs in SC 25 but when normalized to SSN, the halo CME abundance decreases by 7%. Thus, the lower halo abundance confirms that SC 25 is stronger than SC 24. Cumulative distributions of CME speeds (e.g., Gopalswamy 2018) indicates that FW CMEs are responsible for many of the heliospheric consequences of CMEs. In Table 1, the number of FW CMEs in SC 25 is 268, much larger than what was observed in SC 24 (149) over the same epoch. This corresponds to an 80% increase in SC 25. When normalized to SSN, the number of FW CMEs is still higher by ≈ 29%. DH type II bursts are low-frequency radio bursts produced by electrons accelerated in CME-driven shocks. The underlying CMEs are known to be FW (Gopalswamy et al. 2001; 2019b). Not surprisingly, the number of DH type II bursts more than doubled in SC 25, reflecting the enhanced number of FW CMEs in the cycle. When normalized to SSN, the increase in the number of DH type II bursts in SC 25 by ≈ 33% similar to the increase observed in the number of FW CMEs. These increases are over and above that in SSN potentially indicating some non-spot eruptions. Interestingly, the ratio of the number of DH type II bursts to that of FW CMEs remains the same in the two cycles (66% in SC 24 and 68% in SC 25) again indicating a close physical relationship between the two phenomena.

One of the consequences of FW CMEs heading toward Earth is geomagnetic storms. The number of intense geomagnetic storms (Dst ≤ -100 nT) in SC 25 is 18 compared to 12 in SC 24 during the first 61 months in each cycle. The increase also holds when normalized to SSN by 6%. The number of large SEP events increased only slightly in SC 25 but when normalized to SSN, it shows a slight (≈ 17%) decline. The number of GLE events quadrupled in SC 25 and the normalized number shows an increase of 150%. The detection of SEP events and GLE events is affected by their magnetic connection to the observer (GOES and ground-based neutron monitors). The other aspect of solar eruptions, viz., major soft X-ray flares with intensity ≥M1.0 also points to a stronger SC 25: the number of major flares quadrupled relative to SC 24. When normalized to SSN, the number of major flares is still higher by ≈ 182%. While most of the numbers in Table 1 definitely indicate a stronger SC 25, the number of SGRE events shows a significant (≈ 41%) decline from 27 in SC 24 to just 16 in SC 25; the decline is ≈ 57% when normalized to SSN. We expect the number of SGREs to roughly follow the solar cycle variation of the number of FW CMEs and DH type II bursts. The number of SGRE events normalized to the number of DH type II bursts (FW CMEs) is 0.27 (0.18) in SC 24, whereas it drops to 0.09 (0.06) in SC 25. The drop is most likely due to the reduction in LAT's Sun exposure due to the SADA problem.

**Table 2.** SC 24 SGRE events (marked as γ) with the associated SEP events, type II bursts, flares, and CMEs

| No. | Date | γ Start UT | γ end UT | γ Dur [h] | SEP Start UT | SEP Pk [pfu] | DH II UT[a] | m II UT[a] | Flare Location | Flare Class | Flare UT | $V_{CME}$ [km s$^{-1}$] | CME Width |
|---|---|---|---|---|---|---|---|---|---|---|---|---|---|
| 1 | 2011/03/07 | 20:12 | 17:13:15 | 21.02 | 21:45 | 50 | 20:00 | 19:54 | N30W48 | M3.7 | 19:43 | 2125 | H |
| 2 | 2011/06/02 | 07:46 | 14:44:45 | 6.98 | ---- | ---- | 08:00 | ---- | S19E25 | C3.7 | 07:22 | 976 | H |
| 3 | 2011/06/07 | 06:41 | 09:46:00 | 3.08 | 07:20 | 72 | 06:45 | 06:25 | S21W54 | M2.5 | 06:16 | 1255 | H |
| 4 | 2011/08/04 | 03:57 | 06:04:45 | 2.13 | 04:30 | 96 | 04:15 | 03:54 | N19W36 | M9.3 | 03:41 | 1315 | H |
| 5 | 2011/08/09 | 08:05 | 08:40:30 | 0.59 | 08:20 | 26 | 08:20 | 08:01 | N17W69 | X6.9 | 07:48 | 1610 | H |
| 6 | 2011/09/06 | 22:20 | 03:20:30[e] | 5.01 | 22:20 | 9 | 22:30 | 22:19 | N14W18 | X2.1 | 22:12 | 575 | H |
| 7 | 2011/09/07 | 22:38 | 00:42:30 | 2.08 | ---- | HiBg | ---- | 22:38 | N14W28 | X1.8 | 22:32 | 792 | PH |
| 8 | 2011/09/24 | 09:40 | 10:17:00 | 0.62 | ---- | ---- | ---- | 09:35 | N12E60 | X1.9 | 09:21 | 1936 | PH |
| 9 | 2012/01/23 | 03:59 | 19:24:45 | 15.43 | 04:45 | 6310 | 04:00 | ---- | N28W21 | M8.7 | 03:38 | 2175 | H |
| 10 | 2012/01/27 | 18:37 | 22:12:30 | 3.59 | 18:55 | 795 | 18:30 | 18:10 | N27W78 | X1.7 | 17:37 | 2508 | H |
| 11 | 2012/03/05 | 04:09 | 08:24:15 | 4.25 | ---- | ---- | 04:00 | ---- | N17E52 | X1.1 | 03:17 | 1531 | H |
| 12 | 2012/03/07 | 00:24 | 20:52:15 | 20.47 | 02:50 | 6530 | 00:35 | 00:17 | N17E27 | X5.4 | 00:02 | 2684[c] | H |
| 13 | 2012/03/09 | 03:53 | 12:43:45 | 8.85 | ---- | HiBg | 04:10 | 03:43 | N15W03 | M6.3 | 03:22 | 950 | H |
| 14 | 2012/03/10 | 17:44 | 05:21:00 | 11.62 | ---- | HiBg | 17:55 | ---- | N17W24 | M8.4 | 17:10 | 1296 | H |
| 15 | 2012/05/17[d] | 01:47 | 04:52:00 | 3.08 | 01:55 | 255 | 01:40 | 01:31 | N11W76 | M5.1 | 01:25 | 1582 | H |
| 16 | 2012/06/03 | 17:55 | 19:29:30 | 1.57 | ---- | ---- | 18:00 | 17:53 | N16E38 | M3.3 | 17:48 | 605 | PH |
| 17 | 2012/07/06 | 23:08 | 00:33:15[e] | 1.42 | 00:05[h] | 25 | 23:10 | 23:09 | S13W59 | X1.1 | 23:01 | 1828 | H |
| 18 | 2012/10/23 | 03:17 | 06:50:45 | 3.56 | ---- | ---- | ---- | 03:17 | S13E60 | X1.8 | 03:13 | 243 | NH |
| 19 | 2012/11/27 | 15:57 | 17:44:45 | 1.80 | ---- | ---- | ---- | ---- | N05W73 | M1.6 | 15:52 | ---- | -- |
| 20 | 2013/04/11 | 07:16 | 09:39:30 | 2.39 | 08:25 | 114 | 07:10 | 07:02 | N09E12 | M6.5 | 06:55 | 861 | H |
| 21 | 2013/05/13 | 02:17 | 09:45:30 | 7.47 | ---- | ---- | 02:19A | 02:10 | N11E90 | X1.7 | 01:53 | 1270 | H |
| 22 | 2013/05/13 | 16:05 | 00:46:30[e] | 8.69 | 18:00 | 1 | 16:08 | 15:57 | N11E85 | X2.8 | 15:48 | 1850 | H |
| 23 | 2013/05/14 | 01:11 | 07:10:30 | 5.99 | 03:30 | HiBg | 01:16 | 01:07 | N08E77 | X3.2 | 00:00 | 2625 | H |
| 24 | 2013/05/15 | 01:48 | 05:24:30 | 3.61 | 06:35 | 42 | 01:49 | 01:37 | N12E64 | X1.2 | 01:25 | 1366 | H |
| 25 | 2013/10/11 | 07:25 | 08:55:45 | 1.51 | ---- | ---- | 07:23 | 07:11 | N21E103 | M1.5[b] | 07:01 | 1200 | H |
| 26 | 2013/10/25 | 08:01 | 09:28:15 | 1.45 | ---- | ---- | 08:16A | 07:59 | S08E71 | X1.7 | 07:53 | 587 | H |
| 27 | 2013/10/28 | 15:15 | 16:44:15 | 1.49 | 16:00 | HiBg | 15:25 | 15:10 | S06E28 | M4.4 | 15:07 | 812 | H |
| 28 | 2014/01/06 | 07:55 | 09:01:15 | 1.10 | 08:15 | 42 | 07:58 | 07:45 | S15W112 | X3.5 | 07:24 | 1402 | H |
| 29 | 2014/01/07 | 18:32 | 21:28:15 | 2.94 | 19:55 | 1026 | 18:27 | 18:17 | S15W11 | X1.2 | 18:04 | 1830 | H |
| 30 | 2014/02/25 | 00:49 | 09:16:45 | 8.46 | 03:50 | 24 | 00:56 | 00:56 | S12E82 | X4.9 | 00:39 | 2147 | H |
| 31 | 2014/06/10 | 12:52 | 14:59:15 | 2.12 | ---- | ---- | 12:58 | 12:56C | S17E82 | X1.5 | 12:36 | 1469 | H |
| 32 | 2014/06/11 | 09:06 | 10:06:30 | 1.01 | ---- | ---- | ---- | ---- | S18E65 | X1.0 | 08:59 | 829 | NH |
| 33 | 2014/09/01 | 11:11 | 15:06:30 | 3.92 | ---- | ---- | 11:13 | 11:12 | N14E127 | X2.4[b] | 11:05 | 1901 | H |
| 34 | 2014/09/10 | 17:45 | 20:03:45 | 2.31 | 21:35 | 126 | 17:45 | 17:27 | N14E02 | X1.6 | 17:21 | 1267 | H |
| 35 | 2015/06/21 | 02:36 | 16:39:30 | 14.06 | 04:05 | 1066 | 02:33 | 02:25 | N12E16 | M2.6 | 02:03 | 1366 | H |
| 36 | 2015/06/25 | 08:16 | 11:22:15 | 3.10 | 10:05 | 22 | 08:35 | 08:16 | N09W42 | M7.9 | 08:02 | 1627 | H |
| 37 | 2017/09/06 | 12:02 | 06:27:45[e] | 18.43 | 12:35 | 844 | 12:05 | 12:02 | S08W33 | X9.3 | 11:53 | 1571 | H |
| 38 | 2017/09/10[d] | 16:06 | 07:17:00[e] | 15.18 | 16:25 | 1490 | 16:05A | 15:53 | S09W96 | X8.2 | 15:35 | 3163 | H |

[a]A in DH type II column and C in metric type II (m II) column indicate that the onset times were obtained from STEREO-A and Compound Astronomical Low-frequency Low-cost Instrument for Spectroscopy and Transportable Observatory (CALLISTO) dynamic spectra, respectively.

[b]Flare class estimated from EUV data. [c]The 2684 km $s^{-1}$ CME was immediately followed by another halo CME (1825 km $s^{-1}$) from the same active region, accompanied by an X1.3 flare at 01:05 UT. Both eruptions were associated with a single > 100 MeV LAT SGRE event. [d]GLE event. [e]Time corresponds to the next day. "----" in SEP start and SEP peak columns denotes no SEP event; HiBg in SEP peak column means in high background. Note that there are 27 SGRE events during the first 61 months, while there are 38 events over the entire SC 24.

Table 3. SC 25 SGRE events and the associated SEP Events, Type II Bursts, Flares, and CMEs

| No | DATE | γ Start UT | γ end UT | γ Dur [h] | SEP Start UT | SEP Pk [pfu] | DH II UT[a] | m II UT[a] | Flare Location | Flare Class | Flare UT | CME V [km $s^{-1}$] | CME Width |
|---|---|---|---|---|---|---|---|---|---|---|---|---|---|
| 1 | 2021/07/17 | 05:07 | 06:50:30 | 1.72 | ---- | ---- | 05:20A | 05:10C | S20E140 | M5.0 | 04:50 | 1228 | H |
| 2 | 2021/09/17 | 04:20 | 05:37:15 | 1.29 | ---- | ---- | 04:40 | 04:17C | S30E100 | X1.9[b] | 04:17[b] | 1370 | PH |
| 3 | 2021/10/28[c] | 15:35 | 23:28:15 | 7.89 | 16:05 | 29 | 15:37 | 15:29 | S28W01 | X1.0 | 15:17 | 1519 | H |
| 4 | 2022/01/18 | 17:44 | 19:33:30 | 1.82 | ---- | ---- | 17:57 | 17:31 | N18W54 | M1.5 | 17:01 | 1014 | PH |
| 5 | 2022/01/20 | 06:01 | 08:07:45 | 2.11 | 06:25 | 23 | 06:02 | 05:57 | N08W68 | M5.5 | 05:41 | 1431 | PH |
| 6 | 2022/09/29[d] | 12:01 | 12:42:00 | 0.68 | ---- | ---- | 12:05 | 12:00 | N26E106 | X2.0[e] | 11:50[f] | 416[g] | NH |
| 7 | 2022/10/02 | 20:25 | 23:03:15 | 2.64 | 21:00? | <1 | 20:27 | 20:24 | N18W50 | X1.0 | 19:53 | 1086 | H |
| 8 | 2023/12/31 | 21:55 | 01:37:15[h] | 3.70 | 00:15[g] | 20 | 21.57 | 21:46C | N04E73 | X5.0 | 21:36 | 2852 | H |
| 9 | 2024/02/09 | 13:14 | 19:05:15 | 5.85 | 14:00 | 187 | 13:17 | 13:12 | S37W98 | X3.3 | 12:53 | 2782 | H |
| 10 | 2024/02/14 | 04:05 | 05:06:00 | 1.02 | ---- | ---- | 04:16 | 03:55 | S36W160 | ??? | 03:55[i] | 2191 | H |
| 11 | 2024/02/16 | 06:53 | 07:44:15 | 0.85 | 07:15 | 1 | 07:07 | 06:53 | S19W86 | X2.5 | 06:42 | 617 | PH |
| 12 | 2024/07/16 | 13:26 | 21:24:00 | 9.97 | ---- | ---- | 13:44 | 13:21 | S06W85 | X1.9 | 13:11 | 580 | PH |
| 13 | 2024/09/09 | 05:12 | 06:56:30 | 1.74 | ---- | ---- | 05:30 | 05:10 | S13E131 | X3.0[j] | 04:56 | 1522 | H |
| 15 | 2024/09/14 | 15:29 | 02:46:30[h] | 11.29 | 17:10 | 34 | 15:34 | 15:26 | S15E56 | X4.5 | 15:13 | 2366 | H |
| 15 | 2024/10/01 | 22:20 | 01:36:15[h] | 3.27 | ---- | No | ---- | 22:17 | S16E17 | X7.1 | 21:58 | 598 | H |
| 16 | 2024/10/24[d] | 03:57 | 08:19:21 | 4.37[k] | 12:00? | 1 | 03:52 | 03:46 | S05E86 | X3.3 | 03:30 | 2385 | H |

[a]A in DH type II column and C in metric type II (m II) column indicate that the time was obtained from STEREO-A and CALLISTO dynamic spectra, respectively. [b]Flare class reported in Pesce-Rollins et al. (2022) and Yashiro et al. (2024) based on EUVI data; flare time is from EUVI images (cadence is ≈ 5 min). [c]GLE event. [d]Events detected using the Maximum Likelihood method but not the Light Bucket method. [e]Equivalent GOES class based on SolO/STIX data reported in Pesce-Rollins et al. (2024). [f]Onset time of the partially occulted GOES flare. [g]The CME impulsively accelerated to ≈ 1900 km $s^{-1}$ early on in STEREO/EUVI and COR1 FOV and then decelerated to 416 km $s^{-1}$ in the LASCO FOV. [h]The time corresponds to the next day. [i]Flare onset is taken as the starting time of DH Type III burst; the flare peak (04:05 UT) is taken as the end of DH type III bursts preceding type II. [j]From Gopalswamy et al. (2025b). [k]The SGRE increase was delayed, similar to the 2013 May 15 event reported by Ajello (2021). ??? in the Flare Class column indicates that the backsided event has unknown flare class. Note that SC 25 is still in progress and this study is restricted to the first 61 months.

Tables 2 and 3 list the SC 24 and SC 25 SGRE events marked as γ. We excluded "prompt" γ-ray events (Ajello et al. 2021), which are thought to be due to protons accelerated in the flare reconnection site. The first column is the serial number (No.) of events. Columns 2–5 give the SGRE date (YYYY/MM/DD format), onset time, end time, and the duration, respectively. SEP properties are in column 6 (start time) and column 7 (peak >10 MeV integral intensity). The onset time of DH type II bursts is listed in column 8, while that of metric type II (m II) bursts is in column 9. Columns 10–12 give information on the associated soft X-ray flares: heliographic coordinates of the eruption (solar source), soft X-ray flare size, and flare onset time. The flare locations are obtained from the NOAA Space Weather Prediction Center (https://www.swpc.noaa.gov/products/solar-and-geophysical-event-reports) or the SolarSoft archive (http://www.lmsal.com/solarsoft/last_events/). Finally, the CME speed in the coronagraph FOV is in column 13, while the CME width information (H stands for halo, PH for partial halo, and NH for non-halo) is in column 14. The non-halos are normal CMEs with width <120º. These two tables compile relevant energetic phenomena associated with SGRE events. Although we list all SGRE events in SC 24 for completeness, we use only those occurring during the first 61 months to match the corresponding epoch in SC 25.

**2.2 Relative Variation of CME Source Latitudes Associated with SGRE and SEP Events**

Even though SEP events are closest to SGRE events (they share the same acceleration source under the shock paradigm), the latter often occur without the former because of the magnetic connectivity requirement for SEPs. Some SEP events occur when the particle background is high due to a preceding eruption and hence may not be identified as a separate event. Such events are denoted by “HiBg” in Tables 2 and 3. Figure 1 shows the source locations of large SEP and SGRE events as a function of time from Tables 2 and 3 and from the SEP catalog (https://cdaw.gsfc.nasa.gov/CME_list/sepe/). There are 16 SGRE events in SC 24 and 11 in SC 25 with no SEP event having ≥10 pfu detected by GOES; however, almost all these SGRE events are associated with FW CMEs and DH type II bursts. From Fig. 1 we see that only 5 of the 35 SEP events (or 14%) have SGRE association in SC 25 as opposed to 11 of the 30 SEP events or (37%) in SC 24. The SEP events include one GLE event in SC 24 and 4 in SC 25. The second GLE event in SC 24 occurred close to the end of the cycle, therefore, is not within the first 61 months. Among the four GLEs in SC 25, only the first one on 2021 October 28 was associated with an SGRE. The second GLE on 2024 May 11 occurred during a Fermi/LAT data gap. The third GLE occurred on 2024 June 8, the only front-side GLE not associated with an SGRE. The lack of SGRE during the 2024 June 8 GLE is puzzling and needs further investigation because the GLE was accompanied by a high-energy (>100 MeV) SEP event and an intense long-duration type II burst, which are typical of an SGRE event. Finally, the last GLE on 2024 November 21 occurred ≈ 20º behind the west limb and was not associated with an SGRE. Figure 1 also makes it clear that the SEP events and SGRE events all occur in the active region belt (±30º in latitudes) where high magnetic energy is generally available to power these energetic events. It must be noted that the SEP events were identified from GOES proton data, and hence are biased to Earth-arriving particles. Particle detectors in other locations with respect to the Sun-Earth line may see a different set of SEP events.

Tables 2 and 3 show that there is a high degree of association between SGRE and type II bursts. In SC 24, 33 of the 38 SGRE events (or 87%) were associated with DH type II bursts with a similar number (32 of 38 or 84%) associated with metric type II bursts. Only two events lacked type II burst association. In SC 25, all SGRE events are associated with metric type II bursts (100%), while 15 of the 16 (or 94%) are associated with DH type II bursts. Given the moderate association between SEP and SGRE events, DH type II bursts can be seen as better indicators of SGRE events. We take DH type II bursts as proxy for high-energy events, because they are most closely associated with SGRE and the underlying CMEs are a subset of FW CMEs.

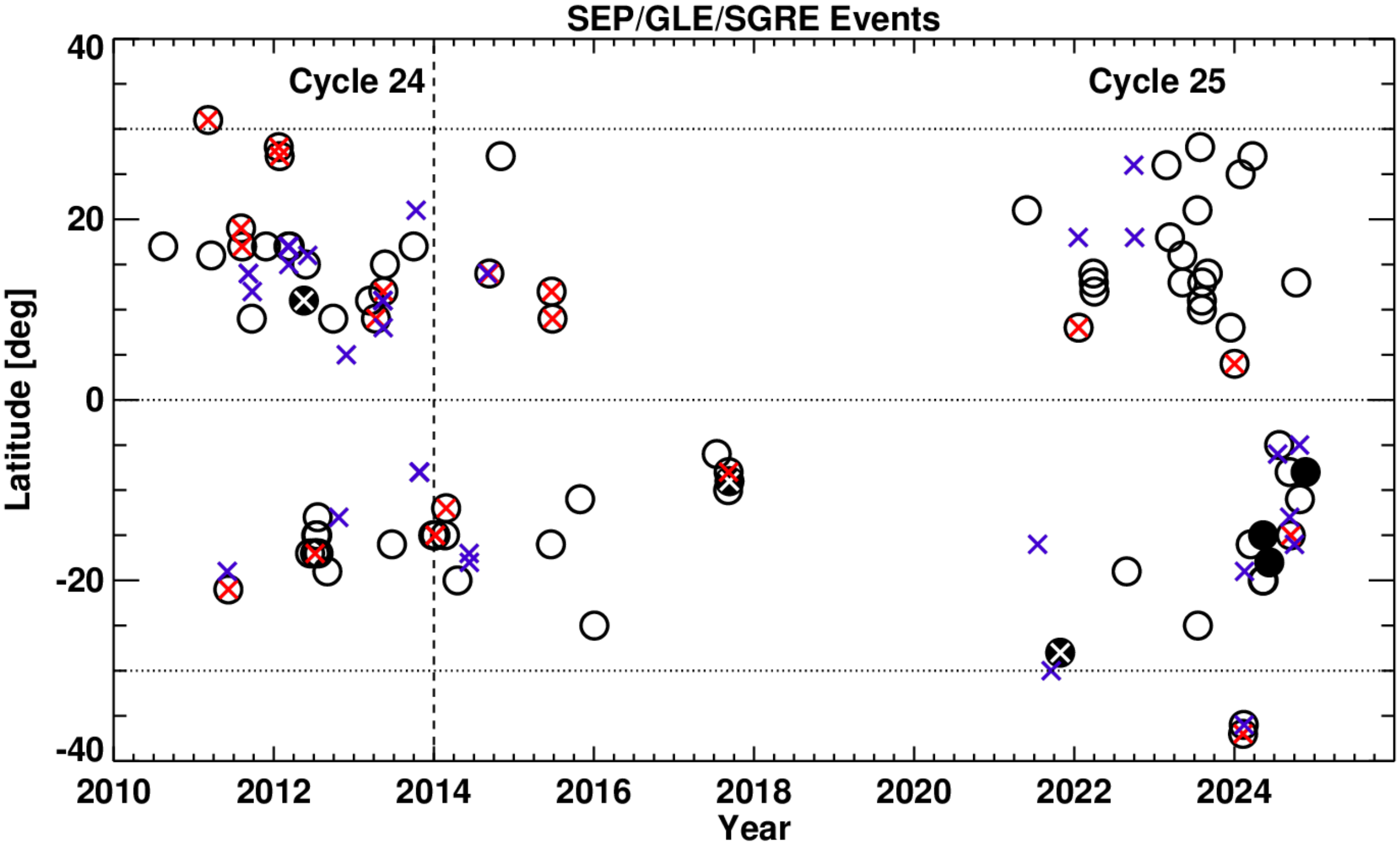

Figure 1. Time-latitude plot of the solar sources of large SEP events and SGRE events in Solar Cycles 24 and 25. SEP events are marked by black circles. Filled black circles represent ground level enhancement (GLE) events in SEPs. SEP events with an SGRE are shown by red x symbols within the black circles. GLEs with SGRE are marked with white x symbols. The blue x symbols denote SGRE events not associated with SEPs. The data points to the left of the vertical dashed line denote events from the first 61 months (December 2008 to December 2013) of Cycle 24 and are compared with the corresponding epoch in Cycle 25 (December 2019 to December 2024). There are 46 data points in SC 24 (30 SEP events with only 11 having SGRE association; 16 SGRE events occurred with no large SEP events detected by GOES). There are also 46 data points in SC 25 (35 SEP events with only 5 of them associated with an SGRE; 11 SGRE events occurred with no large SEP events). The SGRE source locations are in Tables 2 and 4. The SEP source locations are from the SEP catalog (https://cdaw.gsfc.nasa.gov/CME_list/sepe).

The flares involved in SGRE events are major: 23 X-class (61%), 14 M-class (37%) and 1 C-class (2%) flares during entire SC 24 (Table 2); during the first 61 months, the numbers are 14

X-class (51.8%), 12 M-class (44.4%) and 1 C-class (3.7%). In SC 25, the fraction of X-class flares is larger: 12 X-class (80%), 3 M-class (20%) and no C-class (Table 3). All but one SGRE events are associated with white-light CMEs. Out of the 37 SC 24 CMEs in Table 2, 32 are halos (H, 86.5%), 3 partial halos (PH, 8.1%), and 2 non-halos (NH, 5.4%). Limiting to CMEs during the first 61 months in SC 24, we see similar ratios: 22 H out of 26 (or 84.5%), 3 PH (11.5%), and 1 NH (4%). The ratios are slightly different in SC 25: 10 H out of 16 (62.5%), 5 PH (31%) and one NH (6.5%) in SC 25 (Table 3).

One thing obvious from Fig. 1 is the sparse distribution of SGRE events in SC 25. In order to understand the impact of Fermi/LAT data gaps on the SGRE event count in SC 25, we first identify the data gaps, identify the high-energy events during the gaps, and then estimate the number of SGRE events in the cycle. Such an estimate will tell whether the SGRE events are consistent with the fact that SC 25 is slightly stronger than SC 24 as indicated by other energetic events.

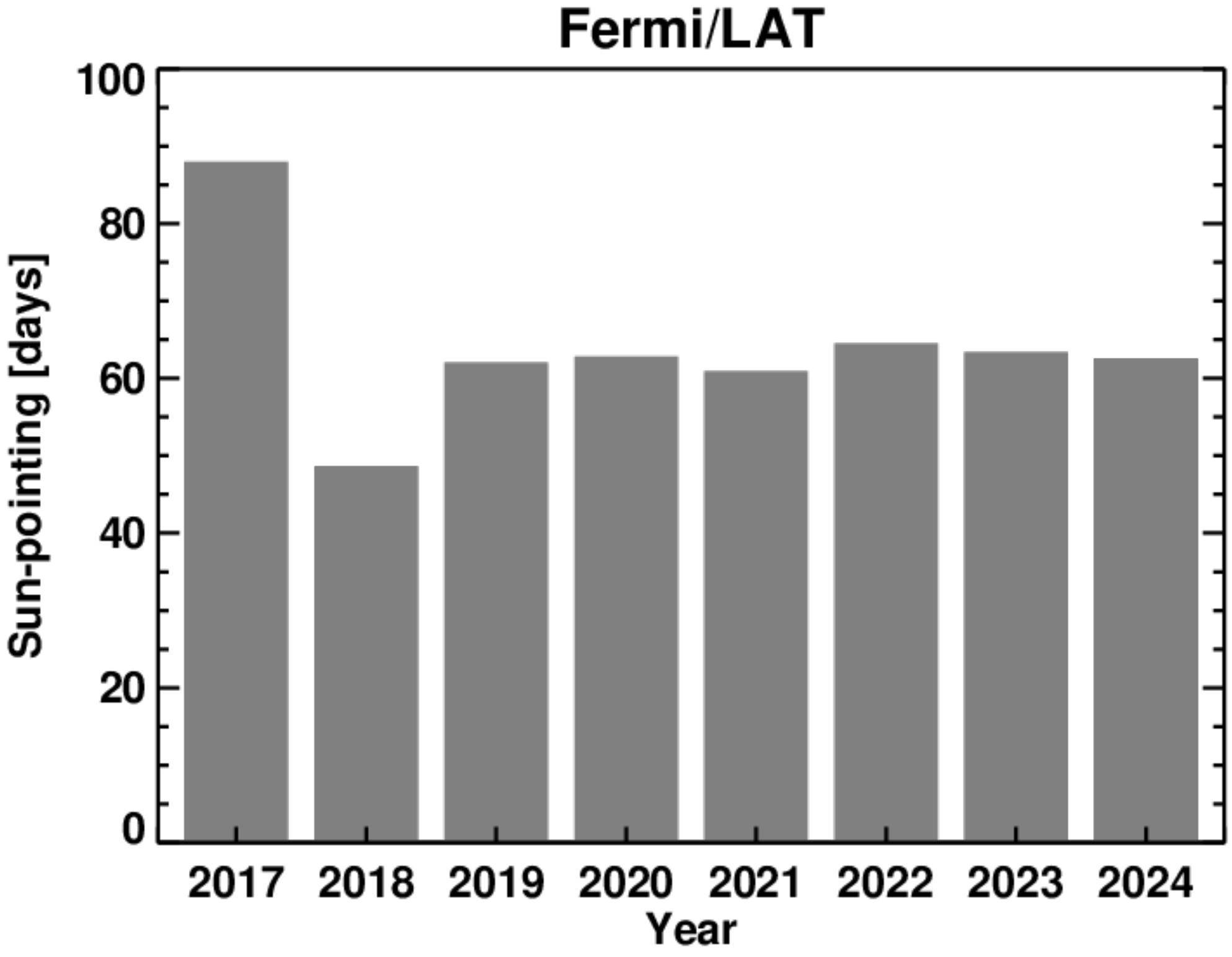

**Figure 2.** Effective solar exposure (number of days) during each year since 2017. The exposure reduction is by about one third after the SADA anomaly that started in March 2018. One can expect a one-third reduction in the number of SGRE events in SC 25 if we assume that the SGRE occurred with the same rate as in SC 24.

### 2.3 Fermi/LAT Data Coverage in SC 25

The reduced solar coverage started on 2018 March 16, towards the end of SC 24. The drive of one of the two solar arrays of the Fermi satellite experienced an anomaly that led to the reduction (https://fermi.gsfc.nasa.gov/ssc/observations/types/post_anomaly/). Although the solar array in question is fully functional, it got stuck at an angle to the LAT boresight because SADA was not able to rotate the array. In order for the non-rotating array to be illuminated, the Sun is kept toward the edge of the LAT FOV resulting in the observational gaps (1–3 weeks at a time) in the Sun exposure. The maximum LAT effective area is reduced to below 4000 $cm^2$ in 2019 and later years compared to a maximum of 5500 $cm^2$ in the pre-2018 period.

Figure 2 shows Fermi/LAT's annual Sun exposure as an effective number of days. The anomaly years (2018 onwards) are compared with one pre-anomaly year (2017). The total annual exposure time during the pre-anomaly years is ≈ 90 days per year determined by the satellite orbital period (≈ 95 min) and the amount of time (≈ 20–30 min per orbit) LAT was pointed to the Sun. The anomaly year 2018 had a larger reduction while redesigning the observing modes (the total annual Sun exposure was only ≈ 50 days). In the years after 2018, the Sun exposure was between 60 and 65 days, amounting to a reduction of ≈ 30%. Lack of LAT observations is substantial, so any SGRE event occurring during these gaps will not be counted. One can infer that more SGRE events might have occurred during the LAT gaps simply based on the fact that there were many FW CMEs, large SEP events, and DH type II radio bursts – the key phenomena associated with SGREs – that occurred during the gaps.

**Table 4.** Fermi/LAT data gaps and the number of energetic events

| Year | No. SGREs | No. Gap days | No. FW CMEs[a] | No. DH Type II[a] | No. SEP[a] |
|---|---|---|---|---|---|
| 2018 | 0 | 132.5 | 0 (0) | 0 (0) | 0 (0) |
| 2019 | 0 | 137.0 | 0 (0) | 1 (0) | 0 (0) |
| 2020 | 0 | 119.5 | 2 (0) | 2 (0) | 0 (0) |
| 2021 | 2 | 130.5 | 11 (1) | 15 (4) | 2 (0) |
| 2022 | 4 | 133.0 | 61 (19) | 34 (8) | 5 (0) |
| 2023 | 1 | 133.5 | 86 (30) | 55 (28) | 12 (7) |
| 2024 | 8 | 133.0 | 108 (35) | 77 (39) | 16 (6) |

[a]Numbers in parentheses are for events occurring during LAT data gaps

Table 4 shows the observed number of SGRE events in SC 25 (column 2) along with the number of Fermi/LAT data gaps in each year (column 3), FW CMEs, DH type II bursts, and large SEP events. The numbers in parentheses correspond to events occurring during Fermi/LAT data gaps. We see substantial numbers of energetic events occurring during the gaps, especially during the solar maximum years (2022–2024). In SC 24, the number of FW CMEs and DH type II bursts is 149 and 99, respectively. The corresponding numbers in SC 25 are 268 and 183 (see Table 1). If the fraction of FW CMEs (18%) and DH type II bursts (27%) remains similar in SC 25, we expect ≈ 48 and 49 SGRE events, respectively. This would imply that 32–33 SGRE events should have occurred during the LAT gaps.

Let us consider the 79 DH type II bursts that occurred during the gaps (the sum of numbers within parentheses in column 5 of Table 4). Recall that only about a quarter of the DH type II bursts are typically associated with SGRE (see Table 1). We need to determine which DH type II bursts are favored to be associated with SGRE events. For this purpose, we make use of the Share et al. (2018) criterion that presence of > 100 keV hard X-ray bursts is necessary for the occurrence of an SGRE event. Fortunately, Fermi/GBM was observing the Sun more or less continuously, so we know how many > 100 keV HXR bursts occurred during LAT data gaps. If a DH type II burst is associated with a > 100 keV HXR burst, it is a good indicator of an SGRE event. Here we confirm that the Share et al. (2018) result on the association of > 100 keV HXR bursts with SGRE events holds good in SC 25 as well.

## 3. SGRE Events and Hard X-ray Bursts

In characterizing the association of HXR bursts with SGRE events, we consider the flux in the Fermi/GBM channel 100–300 keV, which we refer to as a > 100 keV event throughout the paper. While examining the > 100 keV HXR bursts during LAT gaps in SC 25, we found that most of the bursts were of short duration, typically less than a couple of minutes. Some bursts have durations longer than ≈ 5 minutes. Figure 3 shows two GBM HXR bursts identified during two different LAT data gaps. The HXR burst on 2024 October 9 corresponds to an X1.8 GOES flare from the disk center (N13W08) with start, peak, and end times of 01:25 UT, 01:56 UT, and 02:43 UT, respectively. The 100–300 keV count rate gets above the threshold (100 counts/s) around 01:41:37 UT, attains peak (≈ 3252 counts/s) at 01:47:10, and drops below the threshold level at 02:06:04 UT yielding a long duration (LD) of ≈ 24.5 min. The background HXR is the difference between three most sunward detectors and three least sunward detectors. The background level is highly variable, so we chose the 100 counts/s as the threshold level to compute the duration. In some cases, the real background is much lower, so the use of the 100 counts/s threshold underestimates the duration (see later). The HXR burst on 2024 December 21 originates in an impulsive M1.9 GOES flare from S15E62. The SXR flare has a short duration, ≈ 9 min with start, peak, and end times of 00:33 UT, 00:38 UT, and 00:42 UT, respectively. The short duration (SD) HXR burst goes above the 100 counts/s level only for 64 s with its peak (2046 counts/s) coinciding with the SXR peak. The LD burst was associated with a fast (1435 km $s^{-1}$) double-whammy halo CME: it produced a large SEP event with the >10 MeV proton intensity exceeding 1000 pfu and a super-intense geomagnetic storm (Dst ≈ -333 nT). The deprojected speed is ≈ 1857 km $s^{-1}$. The > 100 MeV SEP intensity remained high for more than a day, suggesting that the SEP event is likely to have produced an SGRE event that was not observed due to the LAT gap.

The SD burst was also associated with a narrow (≈ 40º) and slow CME associated with a surge observed by STEREO's Extreme UltraViolet Imager (EUVI, Howard et al. 2008) at 304 Å (00:45 – 00:55 UT). This CME occurred when the SEP background was high due to a previous CME, but SEPs are not expected from such weak and narrow CMEs. The LD event had an intense DH type II burst that lasted at least until 10 UT the next day. The LD HXR burst was associated with a DH type III burst that lasted for ≈ 30 min. The SD HXR was associated with a very short-lived (≈ 1 min) type III burst, but no type II burst. Thus, in all respects, the LD HXR burst is likely to be associated with an SGRE, while it is unlikely that the SD HXR has such an association.

In order to check if SC 25 SGRE events are associated with LD HXR bursts, we examined the > 100 keV Fermi/GBM data corresponding to the LAT events. We specifically use the 100–300 keV channel data to determine the HXR duration and peak flux. The data were obtained with different time resolutions, but we smooth the data over 12 s. When GBM data are interrupted by Fermi night or South Atlantic Anomaly (SAA) passage, we use Konus data when available in estimating the duration. We examined both the trigger mode and waiting mode data (Lysenko et al. 2022) to determine the HXR duration. We use the Konus G2 energy channel (≈ 80–300 keV), which is very similar to the GBM 100–300 keV channel. The Wind spacecraft is located at Sun-Earth L1, so the HXR background is very low and there is no satellite night unlike Fermi. We take the starting (ending) time as the earlier (later) of the GBM or Konus observations. The light travel time is ≈ 5 s from L1 to Earth, so we corrected the Konus time for this.

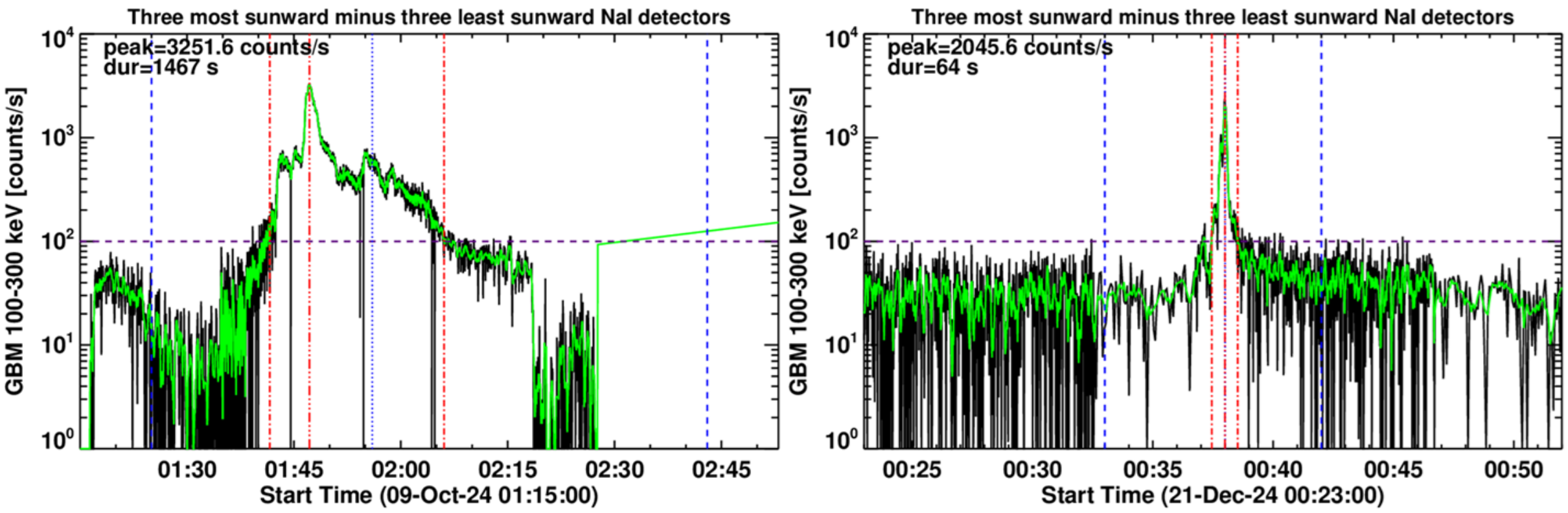

**Figure 3.** Plots of GBM count rates of two > 100 keV hard X-ray bursts (left: 2024 October 9 around 01:15 UT and right: 2024 December 21 around 00:35 UT) in the 100–300 keV energy channel. The bursts occurred during LAT data gaps of 2024 October 8–21 and 2024 December 16–22. The black and green curves are the actual and 12 s averaged count rates. The three vertical red dot-dashed lines in each panel correspond to the onset, peak and end of the HXR burst. The horizontal dashed line at 100 counts/s represents the threshold level we used in identifying HXR bursts. The 100–300 keV HXR durations above 100 counts/s and the peak count rate are noted on the plots. The vertical blue dashed lines mark the start and end of the GOES soft X-ray flare with the vertical blue dotted line showing the GOES SXR peak time.

### 3.1 HXR Bursts Associated with SC 24 SGRE Events

Fermi/LAT observations during most of SC 24 had no major interruptions indicating normal solar coverage until 2018 March 16. The remaining time in SC 24 after the SADA anomaly was towards the end of the cycle, so there was no significant SGRE activity. The SGRE events observed in SC 24 provide a reference to compare those in SC 25 that had reduced solar coverage. Table 5 shows Fermi/GBM HXR bursts associated with SC 24 SGREs along with their durations and peak fluxes (see also Share et al. 2018). For 4 SGRE events, GBM had partial data gaps, so the peak HXR count rate is not available. In these cases, Konus did observe the bursts fully, which we use in the table. We converted the Konus peak count rate (K) in the 80–300 keV channel to an equivalent GBM count rate (G) by establishing a correlation between the fluxes using 14 events in Table 5 observed by both instruments. The regression equation is G = 2.36 K + 929.7 with a correlation coefficient (cc) of 0.97. The Pearson's critical cc for 14 events is 0.780 with a chance coincidence probability $p = 5\times10^{-4}$.

The first three columns of Table 5 are the same as in Table 2, giving the serial number, date, and onset time of SGRE events. For each SGRE event, we give the HXR start time (column 4), end time (column 5), duration above 100 counts/s (column 6), and the peak count rate (column 7). The last two columns give the GOES SXR flare location and start time, respectively for context. The HXR start and end times have a suffix letter G (Fermi/GBM) or K (Wind/Konus) to denote that data of which instrument on the burst was used.

All SGRE events in SC 24 have associated > 100 keV HXR bursts, except for five events: 2011 June 2, 2012 May 17, 2013 October 11, 2013 October 28, and 2014 January 6. The 2013 October 11 event occurred ≈ 13° behind the limb, so it is possible that flare electrons were not able to precipitate on the front side to emit > 100 keV HXR. The 2011 June 2 SGRE event had a GBM data gap and occurred under special circumstances of interacting CMEs (to be discussed later). The flares on 2012 May 17 and 2013 October 28 occurred during Fermi night. The 2014 January 6 was a sub-GLE event that occurred ≈ 22° behind the west limb and the flare impulsive phase occurred during Fermi night. Considering the remaining 33 events, we see that the HXR durations are in the range 2 – 60.35 min. Twenty seven of the 38 events listed in Table 5 occurred during the first 61 months of SC 24. Considering the 23 events with HXR data during the first 61 months, the mean and median durations are 17.96 min and 12.00 min, respectively. If we consider all the events in SC 24, the mean and median durations are not very different: 19.34 min and 17.00 min.

**Table 5.** Hard X-ray burst (>100 keV) information during SC 24 SGRE events

| No. | DATE | SGRE Start UT | HXR Start UT[a] | HXR End UT[a] | HXR Dur [min] | HXR Pk c/s | Flare Location | Flare Class | Flare UT |
|---|---|---|---|---|---|---|---|---|---|
| 1 | 2011/03/07 | 20:12 | 20:02:12G | 20:42:00G | 39.80 | 2120 | N30W48 | M3.7 | 19:43 |
| 2 | 2011/06/02 | 07:46 | DG | DG | DG | DG | S19E25 | C3.7 | 07:22 |
| 3 | 2011/06/07 | 06:41 | 06:24:13G | 06:42:01G | 17.80 | 1453 | S21W54 | M2.5 | 06:16 |
| 4 | 2011/08/04 | 03:57 | 03:48:59G | 04:00:59G | 12.00 | 2271 | N19W36 | M9.3 | 03:41 |
| 5 | 2011/08/09 | 08:05 | 08:01:14G | 08:08:14G | 7.00 | 113214 | N17W69 | X6.9 | 07:48 |
| 6 | 2011/09/06 | 22:20 | 22:17:50G | 22:27:02G | 9.20 | 59017 | N14W18 | X2.1 | 22:12 |
| 7 | 2011/09/07 | 22:38 | 22:35:27G | 22:41:27G | 6.00 | 19018 | N14W28 | X1.8 | 22:32 |
| 8 | 2011/09/24 | 09:40 | 09:35:00G | 09:44:24G | 9.40 | 40038 | N12E60 | X1.9 | 09:21 |
| 9 | 2012/01/23 | 03:59 | 03:52:46G | 04:11:00G | 18.23 | 1444 | N28W21 | M8.7 | 03:38 |
| 10 | 2012/01/27 | 18:37 | 18:09:51G | 18:44:00G | 34.15 | 550 | N27W78 | X1.7 | 17:37 |
| 11 | 2012/03/05 | 04:09 | 03:55:39G | 04:56:00G | 60.35 | 1659 | N17E52 | X1.1 | 03:17 |
| 12 | 2012/03/07 | 00:24 | 00:06:57K | 00:55:00G | 48.05 | 2000 | N17E27 | X5.4 | 00:02[b] |
| 13 | 2012/03/09 | 03:53 | 03:37:50G | 04:03:52G | 26.03 | 363 | N15W03 | M6.3 | 03:22 |
| 14 | 2012/03/10 | 17:44 | 17:40:50G | 18:05:36G | 24.77[c] | 300[c] | N17W24 | M8.4 | 17:10 |
| 15 | 2012/05/17 | 01:47 | DG | DG | DG | DG | N11W76 | M5.1 | 01:25 |
| 16 | 2012/06/03 | 17:55 | 17:53:01G | 17:55:01G | 2.00 | 3313 | N16E38 | M3.3 | 17:48 |
| 17 | 2012/07/06 | 23:08 | 23:00:25G | 23:08:10K | 7.75 | 125584[d] | S13W59 | X1.1 | 23:01 |
| 18 | 2012/10/23 | 03:17 | 03:14:47G | 03:19:47G | 5.00 | 43856 | S13E60 | X1.8 | 03:13 |
| 19 | 2012/11/27 | 15:57 | 15:55:25G | 15:58:01G | 2.60 | 1081 | N05W73 | M1.6 | 15:52 |
| 20 | 2013/04/11 | 07:16 | 07:08:25G | 07:15:48G | 7.38 | 424 | N09E12 | M6.5 | 06:55 |
| 21 | 2013/05/13 | 02:17 | 02:03:13G | 02:10:14G | 7.02[c] | 3000 | N11E90 | X1.7 | 01:53 |
| 22 | 2013/05/13 | 16:05 | 16:04:13G | 16:13:00G | 8.78 | 45142 | N11E85 | X2.8 | 15:48 |
| 23 | 2013/05/14 | 01:11 | 01:02:37G | 01:23:01G | 20.40 | 11346 | N08E77 | X3.2 | 00:00 |
| 24 | 2013/05/15 | 01:48 | 01:36:39G | 01:54:16G | 17.62 | 1936 | N12E64 | X1.2 | 01:25 |
| 25 | 2013/10/11 | 07:25 | No HXR | No HXR | No HXR | No HXR | N21E103 | M1.5 | 07:01 |
| 26 | 2013/10/25 | 08:01 | 07:57:23K | 08:19:00G | 21.78 | 21069[d] | S08E71 | X1.7 | 07:53 |
| 27 | 2013/10/28 | 15:15 | DG | DG | DG | DG | S06E28 | M4.4 | 15:07 |
| 28 | 2014/01/06 | 07:55 | DG | DG | DG | DG | S15W112 | X3.5 | 07:24 |
| 29 | 2014/01/07 | 18:32 | 18:07:19K | 18:11:56K | 4.62 | 2700[d] | S15W11 | X1.2 | 18:04 |
| 30 | 2014/02/25 | 00:49 | 00:42:21G | 01:04:22G | 22.02 | 192537 | S12E82 | X4.9 | 00:39 |
| 31 | 2014/06/10 | 12:52 | 12:42:14G | 12:58:14G | 16.0 | 11361 | S17E82 | X1.5 | 12:36 |
| 32 | 2014/06/11 | 09:06 | 09:02:48G | 09:10:00G | 7.18 | 6663 | S18E65 | X1.0 | 08:59 |
| 33 | 2014/09/01 | 11:11 | 11:04:25G | 11:29:48G | 25.38 | 563 | N14E127 | X2.4 | 11:05 |
| 34 | 2014/09/10 | 17:45 | 17:26:25G | 17:54:25G | 27.98 | 2142 | N14E02 | X1.6 | 17:21 |
| 35 | 2015/06/21 | 02:36 | 02:09:49G | 02:20:00G | 10.18 | 150 | N12E16 | M2.6 | 02:03 |
| 36 | 2015/06/25 | 08:16 | 08:12:24G | 08:29:24G | 17.00 | 30910 | N09W42 | M7.9 | 08:02 |
| 37 | 2017/09/06 | 12:02 | 11:55:29K | 12:40:46G | 45.28 | 113727 | S08W33 | X9.3 | 11:53 |
| 38 | 2017/09/10 | 16:06 | 15:50:11G | 16:39:47G | 49.60 | 125484 | S09W96 | X8.2 | 15:35 |

[a]G and K indicate that the time was determined from GBM and Konus data, respectively.
[b]The X5.4 flare was immediately followed by another X1.3 flare at 01:05 UT. The second flare was also associated with a > 100 keV HXR burst that had a duration of ≈ 23 min (01:05 UT to

01:28 UT) and a peak count rate of 38009 c/s. Both these flares were associated with a single > 100 MeV LAT SGRE event. [c]Duration and peak count rate are lower limits because the start and peak times are unknown. [d]Based on peak count rate from Konus modified using the correlation between peak count rates of SC 24 events observed by both Konus and GBM. [e]Lower limit because the spacecraft night started before the burst ended.

### 3.2 HXR Bursts Associated with SC 25 SGRE Events

Table 6 provides information on > 100 keV HXR bursts associated with the SC 25 SGRE events. Of the 16 SGRE events, only 11 have GBM and/or Konus data as listed in Table 6. The circumstances of the remaining 5 events are as follows. (i) The first SC 25 SGRE event on 2021 July 17 occurred during a Fermi/GBM data gap between July 16 and July 20. Besides, this event occurred ≈ 50° behind the east limb, so it is unlikely that a HXR emission would be detected by GBM or Konus. However, the event was associated with a HXR burst observed by the Spectrometer/Telescope for Imaging X-rays (STIX, Krucker et al. 2020) on board Solar Orbiter (SolO) as reported by Pesce-Rollins et al. (2022). (ii) The 2022 January 18 SGRE had a GBM data gap, and no event was reported by Konus. (iii) The 2024 February 14 event was completely backsided with no HXR detector operating on the backside during the event. (iv) The 2024 September 9 event was also associated with a backside eruption but a HXR burst was detected by SolO (Gopalswamy et al. 2025b). (v) The 2024 October 1 event occurred during Fermi night, but the onset, peak, and early declining phases were observed by Konus, indicating that the HXR burst had a duration of at least ≈ 4.2 min. From the 11 events with known start and end times of > 100 keV HXR bursts, we see that the durations are in the range 4.32–41.37 min with an average (median) value of 16.64 (11.33) min. These numbers are not too different from those in the corresponding epoch of SC 24. Thus, the HXR durations in SGRE events are similar to those of the LD events found during LAT gaps (see Fig. 3 for an example). The association of > 100 keV HXR during SGRE events reported by Share et al. (2018) are thus confirmed with the additional constraint that the HXR duration is ≳ 5 min.

**Table 6.** Hard X-ray burst (>100 keV) information during SC 25 SGRE events

| No. | DATE | SGRE Start UT | HXR Start UT[a] | HXR End UT[a] | HXR Dur [min] | HXR Pk c/s | Flare Location | Flare UT |
|---|---|---|---|---|---|---|---|---|
| 1 | 2021/07/17 | 05:07 | DG | DG | DG | DG | S20E140 | 04:50 |
| 2 | 2021/09/17 | 04:20 | 04:14:00G | 04:20:00G | 6.0 | 450 | S30E100 | 04:17 |
| 3 | 2021/10/28 | 15:35 | 15:27:30K | 16:08:52G | 41.37 | 2514 | S28W01 | 15:17 |
| 4 | 2022/01/18 | 17:44 | DG | DG | DG | DG | N18W54 | 17:01 |
| 5 | 2022/01/20 | 06:01 | 05:54:07G | 06:04:43G | 10.60 | 2161 | N08W68 | 05:41 |
| 6 | 2022/09/29 | 12:01 | 11:54:58K | 11:59:57K | 4.32[b] | 603 | N26E106 | 11:50 |
| 7 | 2022/10/02 | 20:25 | 20:20:13G | 20:30:25G | 10.22 | 9117 | N18W50 | 19:53 |
| 8 | 2023/12/31 | 21:55 | 21:38:44K | 22:08:26G | 29.70 | 2566 | N04E73 | 21:36 |
| 9 | 2024/02/09 | 13:14 | 13:01:44K | 13:13:04K | 11.33 | 5632 | S37W98 | 12:53 |
| 10 | 2024/02/14 | 04:05 | DG | DG | DG | DG | S36W160 | 03:55 |

| 11 | 2024/02/16 | 06:53 | 06:50:55K | 06:59:20G | 8.42 | 12909 | S19W86 | 06:42 |
|---|---|---|---|---|---|---|---|---|
| 12 | 2024/07/16 | 13:26 | 13:18:55K | 13:35:48G | 16.88 | 106 | S06W85 | 13:11 |
| 13 | 2024/09/09 | 05:12 | DG | DG | DG | DG | S13E131 | 04:56 |
| 15 | 2024/09/14 | 15:29 | 15:17:02G | 15:36:15G | 19.22 | 31516 | S15E56 | 15:13 |
| 15 | 2024/10/01 | 22:20 | DG | DG | DG | DG | S16E17 | 21:58 |
| 16 | 2024/10/24 | 03:57 | 03:38:36G | 04:03:37G | 25.02 | 8099 | S05E86 | 03:30 |

[a]G and K indicate that the time was determined from GBM and Konus data, respectively. [b]The time profile is quite flat at the end not returning to the pre-event level, so the duration is likely an underestimate.

### 3.3 HXR Bursts Associated with DH Type II Bursts in LAT Gaps

In this subsection we consider all the DH type II bursts observed during LAT data gaps and their association with > 100 keV HXR bursts from Fermi/GBM and Wind/Konus. We identified 79 DH type II bursts during the LAT gaps. We do know that almost all SGRE events are associated with type II radio bursts, while only ≈ 25% of DH type II bursts are associated with SGRE events (see Table 1). Examining the association of > 100 keV HXR bursts we find that out of the 79 DH type II bursts, only 27 (or 34%) had associated HXR bursts. Out of the remaining 52 DH type II bursts, 26 had no HXR burst association. The remaining 26 DH type II bursts occurred during Fermi night or during South Atlantic Anomaly (SAA) crossing. Table 7 lists the 27 type II bursts with overlapping Fermi/GBM observations. All HXR bursts in Table 7 have long durations except for three events that lasted for 0.77, 2.02, and 2.22 min. These durations are likely underestimates as will be discussed later in this subsection. Most of the flares in Table 7 are of X class, with 12 of them of M-class (or 44%), similar to what was observed during SGREs of SC 24. As in Tables 2 and 3, most of the CMEs are halos (19 out of 27 or 70%), with 5 partial halos (19%), and three non-halos (NH, 11%). Having a large fraction of halos indicates that the CME population underlying DH type II bursts is very energetic (Gopalswamy et al. 2019b).

**Table 7.** List of > 100 keV HXR bursts associated with DH type II bursts in LAT data gaps

| **No** | **DATE** | **STIME** | **ETIME** | **Dur [min]** | **HXR Flux** | **Flare Location** | **Class** | **Flare Time** | **CME V [km s$^{-1}$]** | **CME Width** |
|---|---|---|---|---|---|---|---|---|---|---|
| 1 | 2022/04/29 | 07:20:39K | 07:21:25K | 0.77 | 260 | N25W37 | M1.2 | 07:15 | 1292 | PH |
| 2 | 2023/02/17 | 19:57:39K | 20:24:59G | 27.33 | 1528 | N25E64 | X2.2 | 19:38 | 1315 | H |
| 3 | 2023/05/09 | 18:28:26K | 18:56:23G | 27.95 | 193 | N13W31 | M4.2 | 18:20 | 1209 | H |
| 4 | 2023/06/20 | 16:57:49G | 17:08:59K | 11.17 | 267 | S17E73 | X1.1 | 16:42 | 1113 | H |
| 5 | 2023/07/28 | 15:43:42K | 16:11:50G | 28.13 | 164 | N23W94 | M4.1 | 15:39 | 1896 | H |
| 6 | 2023/08/05 | 22:00:59K | 22:21:01G | 20.03 | 1058 | N11W77 | X1.6 | 21:45 | 1647 | H |
| 7 | 2023/08/07 | 20:37:35K | 21:08:09K | 30.57 | 244 | N13W98 | X1.5 | 20:30 | 1851 | H |
| 8 | 2023/09/19 | 09:32:44K | 09:37:26K | 4.70 | 115 | N07E51 | M1.8 | 09:23 | 418 | PH |
| 9 | 2023/09/19 | 20:08:38G | 20:10:39G | 2.02 | 1107 | N08E45 | M4.0 | 20:01 | 483 | NH |
| 10 | 2023/11/28 | 19:37:03G | 19:50:51G | 13.80 | 536 | S16W00 | M9.8 | 19:35 | 741 | H |
| 11 | 2024/05/03 | 02:18:24G | 02:24:36G | 6.20 | 3123 | N25E07 | X1.6 | 02:11 | 808 | H |
| 12 | 2024/05/08 | 04:46:52G | 05:09:28G | 22.60 | 116 | S22W11 | X1.0 | 04:37 | 530 | H |
| 13 | 2024/05/08 | 11:58:00K | 12:12:58K | 14.97 | 211 | S20W17 | M8.7 | 11:26 | 677 | H |
| 14 | 2024/05/09 | 08:59:39K | 09:45:50G | 46.18 | 308 | S20W26 | X2.2 | 08:45 | 1280 | H |
| 15 | 2024/05/09 | 17:30:37G | 17:35:25G | 4.80 | 363 | S14W28 | X1.1 | 17:23 | 1024 | H |
| 16 | 2024/05/10 | 06:42:02G | 06:51:39G | 9.62 | 12078 | S17W34 | X3.9 | 06:27 | 953 | H |
| 17 | 2024/05/11 | 01:14:24G | 01:37:49G | 23.42 | 56565 | S15W45 | X5.8 | 01:10 | 1614 | H |
| 18 | 2024/05/14 | 02:04:24G | 02:10:36G | 6.20 | 13609 | S19W88 | X1.7 | 02:03 | 881 | NH |
| 19 | 2024/05/14 | 16:46:24G | 16:50:57K | 4.55 | 146483 | S18W96 | X8.7 | 16:46 | 2010 | H |
| 20 | 2024/05/15 | 08:15:47G | 08:37:55G | 22.13 | 46539 | S18W98 | X3.5 | 08:13 | 1648 | H |
| 21 | 2024/07/28 | 10:35:39G | 10:46:26G | 10.78 | 129 | S11W40 | M7.7 | 10:27 | 918 | NH |
| 22 | 2024/08/02 | 04:38:26K | 04:43:15K | 4.82 | 167 | S14W90 | M7.3 | 04:23 | 1141 | PH |
| 23 | 2024/08/05 | 05:20:04K | 05:24:50G | 4.77 | 2097 | S11E62 | M6.1 | 05:36 | 973 | PH |
| 24 | 2024/08/07 | 13:46:11G | 13:48:24G | 2.22 | 1842 | S11E03 | M4.5 | 13:30 | 658 | H |
| 25 | 2024/09/22 | 21:22:38G | 21:32:00K | 9.37 | 141 | S20E63 | M3.7 | 21:12 | 1256 | H |
| 26 | 2024/10/09 | 01:42:27K | 02:18:26G | 35.98 | 3151 | N13W08 | X1.8 | 01:25 | 1435 | H |
| 27 | 2024/10/09 | 15:44:01G | 15:48:25G | 4.40 | 46286 | S10W83 | X1.4 | 15:44 | 874 | PH |

### 3.4 Comparing HXR Properties during SGREs and DH Type II Bursts

We measured the HXR durations of SGREs (during SCs 24 and 25) and LAT-gap DH type II bursts in SC 25. From Tables 2, 3, and 7, we see that the underlying CMEs are energetic, associated with major flares (M and X class), and the HXR bursts have long durations. The average sky-plane speeds are: 1463 km $s^{-1}$ (SC 24), 1497 km $s^{-1}$ (SC 25), and 1141 km $s^{-1}$ (DH type II). SGREs of SC 24 and DH type II bursts have similar fraction of M-class flares, while SC 25 SGREs have mostly X-class flares, although the sample size is small for SGREs in SC 25. Now, we investigate if the > 100 keV association is similar among these populations.

Figure 4a–c compare the > 100 keV HXR duration distributions among: (i) SGRE events in SC 24, (ii) SGRE events in SC 25, and (iii) LAT-gap DH type II bursts. The duration distributions are similar. The median HXR duration for SC 24 SGREs, SC 25 SGREs, and LAT-gap DH type II bursts are 17.00 min, 11.33 min, and 10.78 min, respectively. The mean durations are also similar: 19.34 min (SC 24), 16.64 (SC 25), and 14.8 min (LAT-gap DH type II bursts). Kolmogorov-Smirnov (KS) tests (Kirkman, 1996) comparing these durations taken two (N1, N2) at a time show that these duration distributions are not statistically different. Table 8 presents these comparisons along with the KS test statistic D, which is the maximum difference between the cumulative distributions of a pair compared, and the critical value ($D_c$) of the KS statistic, which needs to be exceeded for the distributions to be statistically different. In all cases in Table 8, $D < D_c$ indicating that there is no significant difference between the distributions compared. In addition to populations (i) – (iii), we also considered another set (i) + (iii) consisting of SC 25 SGREs and LAT-gap DH type II bursts. The HXR durations in this set is similar to the other ones as confirmed by the KS test.

**Table 8.** Comparison of various duration distributions

| **Duration Comparison** | **N1, N2** | **D** | **$D_c$** |
|---|---|---|---|
| SC25 vs. SC 24 | 11, 33 | 0.1515 | 0.4735 |
| SC 25 vs. DHgap | 11, 27 | 0.2424 | 0.4865 |
| DHgap vs. SC 24 | 27, 33 | 0.2559 | 0.3529 |
| DHgap + SC 25 vs. SC 24 | 38, 33 | 0.1906 | 0.3235 |

The good overlap of > 100 keV HXR durations among DH type II bursts and SGRE events in the two cycles is illustrated in Fig. 4d as a scatter plot between the 100–300 keV HXR duration and the peak count rate. The overlap is consistent with the close relation between DH type II bursts and SGRE events. The fact that DH type II data points are indistinguishable from the SGRE data points is important in estimating the number of SGRE events in SC 25.

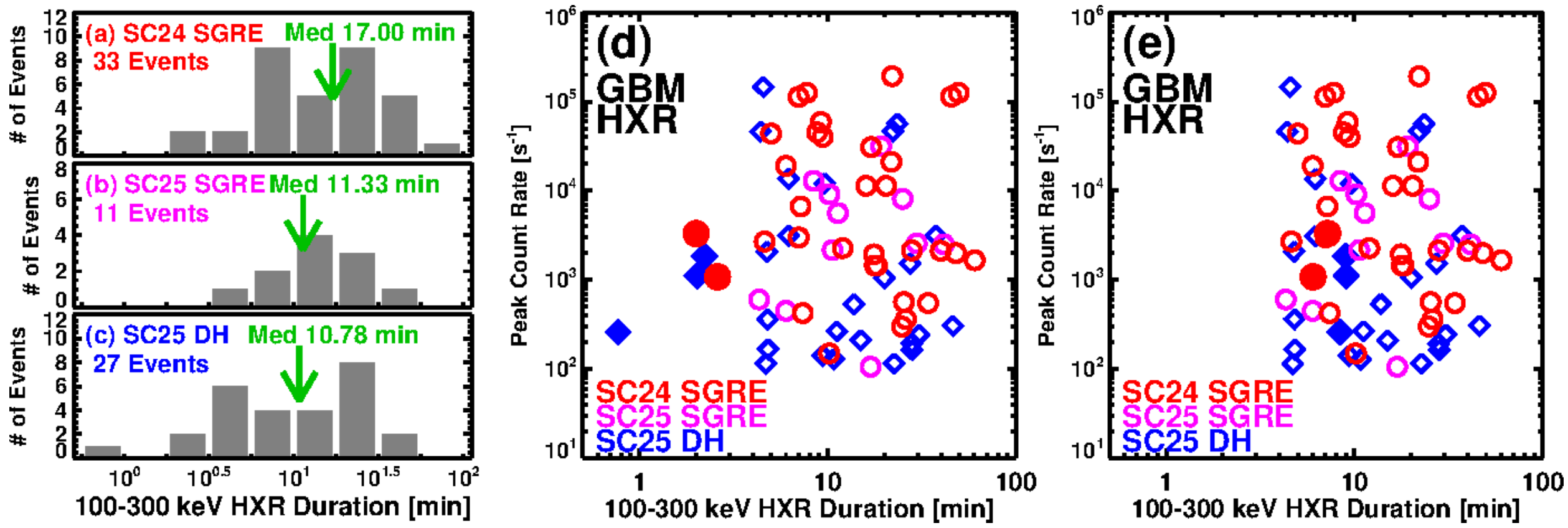

**Figure 4.** Distributions of > 100 keV HXR bursts associated with (a) SC 24 SGREs, (b) SC 25 SGREs, and (c) SC 25 DH type II bursts that occurred during the LAT data gaps. (d) Scatter plot of the > 100 keV HXR bursts vs. their peak count rate for the three populations shown in (a–c). The number of events (N) in each population is shown on the plots. The leftmost five events that have HXR durations < 2.6 min are indicated by filled symbols. The durations of these events are severely underestimated due to the 100 counts/s threshold. (e) Same as (d) but after correcting the durations of the filled symbol events in (d).

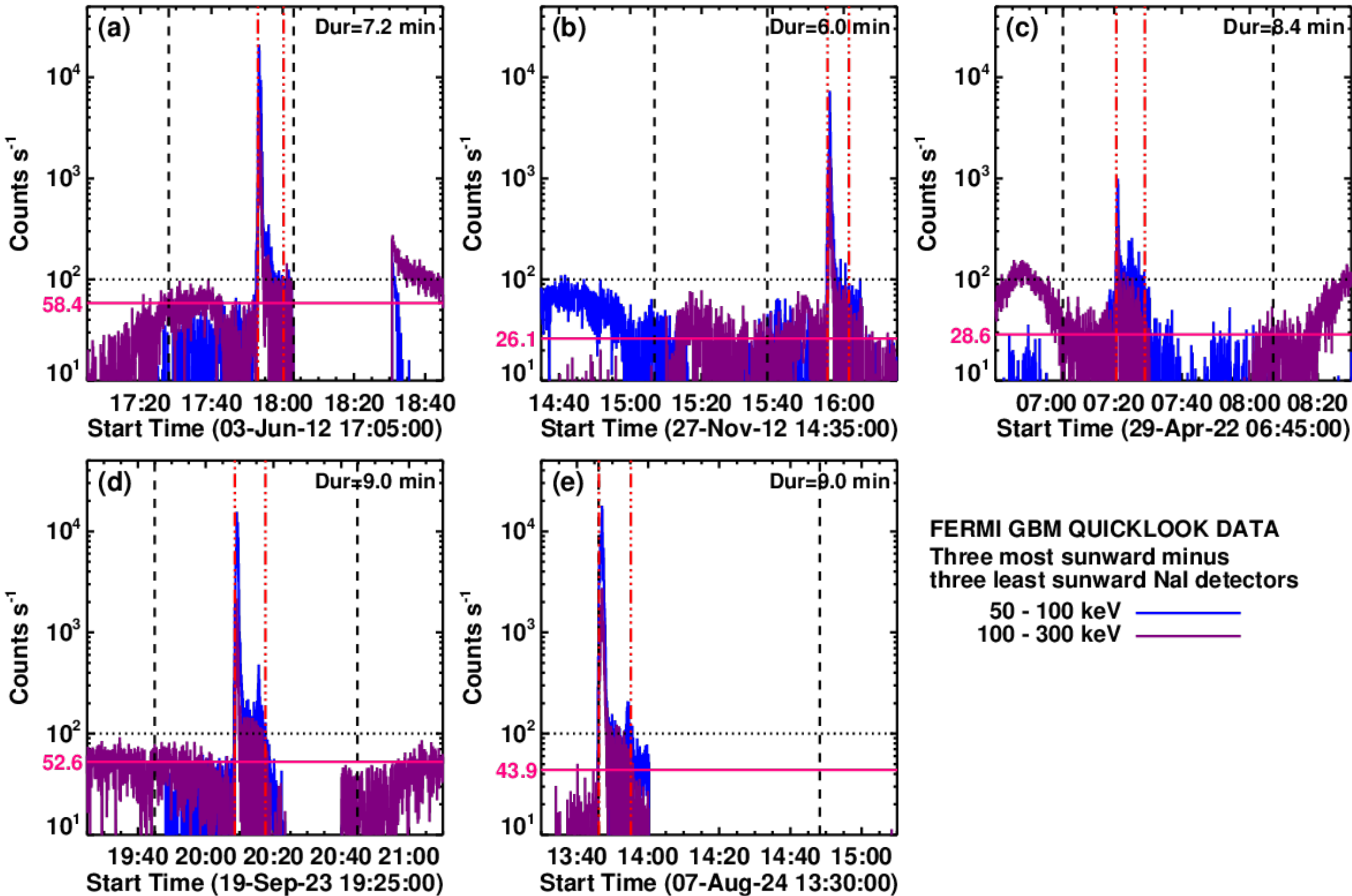

**Figure 5.** Fermi/GBM light curves in the energy channels 50–100 keV (blue) and 100–300 keV (purple) of five HXR bursts that showed the shortest > 100 keV durations. The 2012 June 3 and 2012 November 27 are associated with SC 24 SGREs. The other three are associated with DH type II bursts that occurred during the Fermi/LAT data gap. In all events there is a second HXR peak (prominent in the 50–100 keV channel) within the impulsive phase. The dotted horizontal line at 100 counts/s was originally used to estimate the duration. The actual background is smaller than 100 counts/s by a factor of ≈ 2–4, and is shown by the horizontal red line. The background was determined based on the average counts over 5 minutes before the starting point and after the ending point above 100 counts/s. The greater of the two is used as the background and is given on the plots. The HXR duration is marked by the pair of red dot-dashed vertical lines in each plot. The vertical black dashed lines denote Fermi's day-night boundaries.

There are five HXR bursts of duration in the range 0.77 to 2.60 min shown by filled symbols at the left end of the plot in Fig. 4d. Two of them are SC 24 SGRE events (Table 5: Nos. 16, 19) and three are LAT-gap DH type II bursts (Table 7: Nos. 1, 9, 24). A closer look at the GBM light curves indicates that the background HXR level in these events is much lower than the 100 counts/s that we used as the threshold for measuring the duration. When the actual background is used, the durations of these five events are significantly longer. Figure 5 shows the light curves of these events indicating that their estimated durations range from 6 min to 9 min, on par with the other HXR events in Tables 5–7. The scatter plot in Fig. 4e uses the corrected durations for these five events. Using the actual background allows the inclusion of secondary HXR peaks in the impulsive phase as can be seen in Fig. 5. These considerations suggest that the association of

> 100 keV HXR bursts of long duration (≳ 5 min) along with DH type II bursts is a necessary condition for the production of an SGRE. This result is consistent with the fact that only long-duration > 100 keV HXR bursts are associated with CMEs, while the impulsive short duration ones are not associated with CMEs (Mäkelä et al. 2026, under preparation). These authors investigated all the > 100 keV GBM HXR bursts that occurred after the SADA anomaly (2018–2024). A total of ≈ 200 bursts were identified including those in Tables 5 and 6. Only 137 of the 200 HXR bursts are associated with CMEs. When not associated with a CME, the HXR durations are much smaller, typically <1 min.

### 3.5 Estimating the Number of SGRE Events in SC 25

The above discussion shows that SGRE events and the 27 LAT-gap DH type II bursts have similar characteristics in terms of CME speeds, flare sizes, and particularly > 100 keV HXR durations. Therefore, we suggest that these LAT-gap type II bursts must be indicative of SGRE. Adding these 27 to the 16 SGRE events that were actually observed by Fermi/LAT brings the total number to 43 in SC 25. A second way is to assume that the number of SGRE events follows SSN: the average SSN increased from 56.9 in SC 24 to 79.0 in SC 25, or by 39%, so the number in SC 25 should be ≈ 38 (27×1.39). However, the number of FW CMEs and DH type II bursts increased more than SSN did by 29% and 33%, respectively. Therefore, the SSN-based estimate needs to be boosted by 29% and 33%, respectively to yield 48 (27×1.39×1.29) and 50 (27×1.39×1.33) SGRE events in SC 25. A third way is to use the observation that 18% of all FW CMEs and 27% of all DH type II bursts produced SGRE events in SC 24 (see Table 1). If these association rates hold good in SC 25, we expect ≈ 48 (18% of 268 FW CMEs) and 49 (27% of 183 DH type II bursts) SGRE events. These numbers are very similar to the SSN-based estimates and are only slightly higher than the estimate based on HXR burst association. The average value from the three methods is 46 SGRE events. Given the variation in the heliospheric properties that modulate the shock strength, we think that the estimated number of SGRE events in SC 25 is reasonable and consistent among different methods.

## 4. Discussion

Fermi/LAT > 100 MeV, quick-look plots consist of one data point per solar exposure obtained every 1.5 hours over an exposure time of 20–40 minutes. The > 100 MeV flux is calculated by two different methods: the Light Bucket and the Maximum Likelihood methods (https://hesperia.gsfc.nasa.gov/fermi/lat/qlook/LAT_qlook_plots.htm). In our examination of the quick-look Fermi/LAT 4-day plots, we found a few events that are present in the Maximum Likelihood plots, but not in the Ligh Bucket plots as in events Nos. 6 and 16 in Table 3. If any of these events can be confirmed as SGRE events, the current estimate of the number (43) will be revised upwards moving it closer to the estimates from other methods.

It has been known for a long time that gradual HXR (GHX) bursts are closely associated with large SEP events (Kiplinger 1995; Garcia 1994). GHX typically have durations > 10 min as opposed to impulsive HXR (IHX) bursts that have a duration <2 min. The long duration GBM HXR bursts in our study clearly belong to the GHX bursts from the past. One of the main characteristics of GHX bursts is that the associated flares have a lower temperature, which is proportional to the soft X-ray flux ratio R between the shorter (0.04–0.5 nm) and longer (0.1–0.8 nm) wavelength channels. Ling and Kahler (2020) and Kahler and Ling (2020) have shown that R is also a good indicator of FW CMEs. Large SEP events and DH type II bursts are also associated with FW CMEs because these CMEs drive shocks that accelerate electrons and protons (Gopalswamy et al. 2001; 2004). Thus, we see that the GHX link to SEPs and FW CMEs extends to DH type II bursts, which we utilized in this paper. Another key characteristic of GHX bursts is their soft-hard-harder spectral profile (Kiplinger 1995). As noted by Share et al. (2018), > 100 keV HXR bursts have a harder spectrum than the ones at lower energies (< 50 keV). We see that all phenomena associated with energetic CMEs (large SEP events, DH type II bursts, and shocks) are accompanied by > 100 keV HXR bursts. Although there is no definite answer for the observed connection between > 100 keV HXR bursts and FW CMEs, they are inevitably tied together by the same magnetic reconnection process. While both DH type II bursts and SEPs are definitely CME related (via shock), SEPs have an additional potential source-flare reconnection site. Complex type III bursts that invariably accompany large eruptions are indicative of open magnetic field lines along which electrons accelerated from the flare site escape into the interplanetary medium (Cane, Erickson, and Prestage 2002, Gopalswamy et al. 2012, Winter and Ledbetter, 2015; Gopalswamy et al. 2023b). This escape channel should be available for flare ions as well that may be reaccelerated by the accompanying CME-driven shock (Share et al. 2018).

The 2011 June 2 event did not have GBM observations, but the underlying CME was fast and (976 km $s^{-1}$) wide (halo CME) with an associated DH type II burst. Share et al. (2018) investigated this event and pointed out that the flare was observed by the Reuven Ramaty High-Energy Solar Spectroscopic Imager (RHESSI, Lin et al. 2002) in hard X-rays only for the first three minutes (07:33 UT to 07:36 UT). So, we cannot say whether the HXR emission was long duration or not. Furthermore, this event had a special situation in that there were two CMEs launched from the same active region separated by only ≈ 1 hr. The associated flares were weak with soft X-ray class of C1.4 and C3.7. Figure 6 shows two prominence eruptions (P1, P2) associated with the two soft X-ray flares (F1 and F2) and homologous EUV waves (Wave 1 and Wave 2). The SOHO/LASCO CME catalog lists a slow (253 km $s^{-1}$) but wide ($61^{o}$) CME appearing at 07:24 UT from the first eruption and a fast (976 km $s^{-1}$) halo CME with a first-appearance time of 08:12 UT in the second eruption. Given the disk-center location of the source active region (AR 11227), these sky-plane speeds are expected to be much smaller than the true speeds. The STEREO-Behind spacecraft was located at E93 at the time of the eruptions, so the two CMEs were observed as limb events in its FOV (at W73 and W78). The first appearance time and height of the of the CMEs are 06:45:34 UT at 1.73 Rs (CME 1) and 07:45:35 UT at 2.34 Rs (CME 2). Tracking the leading edges, we found that both CMEs were accelerating in the COR1 FOV and had a speed of 818 km $s^{-1}$ (CME 1) and 1237 km $s^{-1}$ (CME 2) by the time they reached the edge of the FOV around 4 Rs. Gopalswamy et al. (2019a) suggested that the first CME might have provided a mirroring situation, so that the protons accelerated by the second CME were directed sunward to precipitate and produce SGRE.

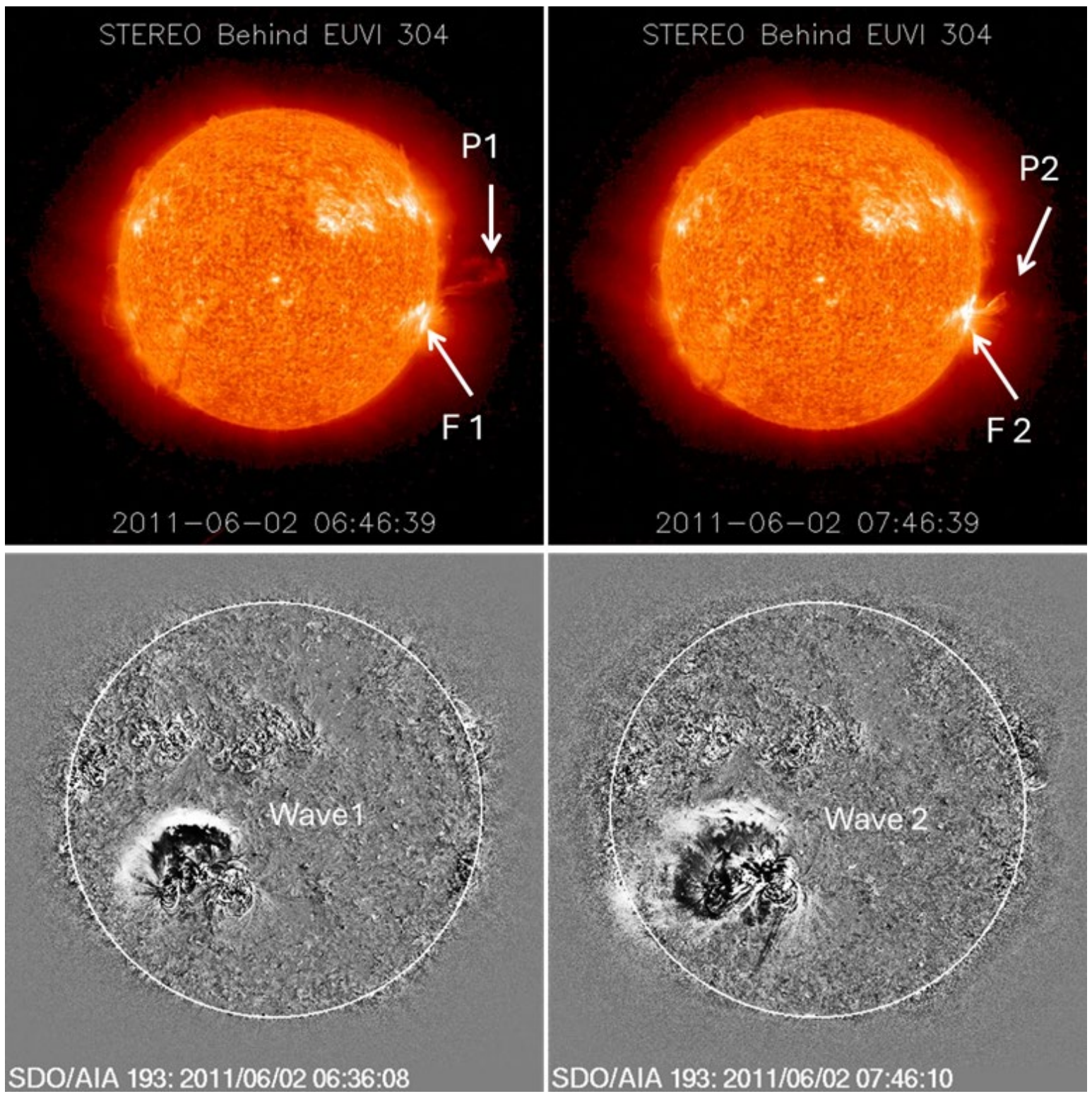

**Figure 6.** Two eruptions from NOAA active region 11227 that occurred during 2011 June 2. (top) Prominence eruptions (P1 and P2) and flares (F1 and F2) associated with the event in STEREO-Behind EUV images at 304 Å taken at 06:46 UT and 07:46 UT. In soft X-rays (GOES), F1 was a C1.4 flare from S19E20 (06:30 UT) while F2 was a C3.7 flare from S19E25 (07:22 UT). (bottom) Homologous EUV disturbance (Wave 1 and Wave 2) revealed by the 193 Å running difference images at 06:36 UT and 07:46 UT from the Atmospheric Imaging Assembly (Lemen et al. 2012) on board the Solar Dynamics Observatory (SDO/AIA).

## 5. Summary

Studies of Fermi/LAT observations have shown that there is a close relation among SGRE events, FW CMEs, large SEP events, and interplanetary type II bursts. The reason behind this relation is as follows: the FW CMEs drive strong shocks that accelerate protons responsible for SGRE events and SEP events, while the accelerated electrons result in type II radio bursts. The energetic ions responsible for SGRE are thought to propagate from the shock toward the Sun while those propagating away into space are detected as SEP events. Previous studies have shown that the solar cycle variation of the occurrence rates of SEP events and IP type II bursts generally follow that of the FW CMEs. Since SGRE events are associated with these energetic phenomena, we expected their occurrence rate to have a similar behavior. However, we observed a reduction in the number of SGRE events in SC 25 relative to the weaker SC 24. The reduction seems to be primarily due to Fermi/LAT's reduced Sun exposure following the malfunction of the solar array drive assembly since 2018 March 16. Invoking the close connection between > 100 keV hard X-ray bursts and SGRE events, we identified such bursts during Fermi/LAT data gaps. When such hard X-ray bursts are also associated with DH type II bursts in the gaps, we suggest that such combination is indicative of an SGRE event. By this method, we estimated that the total number of SGRE events in SC 25 is ≈ 43. If the association rate of SGRE events with FW CMEs and type II bursts observed in SC 24 holds good, we expect 48 and 49 SGRE events in SC 25. If the solar cycle variation of SGRE events is similar to other energetic events, we estimate 48 (FW CMEs) and 50 (type II bursts) SGRE events in SC 25. The estimated number of SGRE events is consistent with the fact that SC 25 is stronger than SC 24. The primary conclusions of this work are as follows.

1. The halo CME abundance in SC 25 is lower than that in SC 24 consistent with the previous studies which showed that weaker solar cycles have a higher halo abundance.
2. Energetic phenomena such as fast and wide CMEs, halo CMEs, type II radio bursts, GLE events, and intense geomagnetic storms all indicate a stronger SC 25.
3. Although the number of large SEP events was higher in SC 25, the increase was not commensurate with solar activity (SSN). The change in the ambient magnetosonic speed and the magnetic connectivity might have affected the SEP event number.
4. The number of fast and wide CMEs and DH type II bursts increased in SC 25 more than SSN did by 29% and 33%, respectively.
5. While the association rate of SGRE events with SEP events is low due to magnetic connectivity, it is very high (> 90%) with type II bursts (metric and/or DH) since there is no connectivity issue for the bursts.
6. The ratio of FW CME to DH type II numbers remains the same between SCs 24 (66%) and 25 (68%) confirming their physical connection (fast and wide CMEs drive shocks that accelerate electrons to produce type II bursts).
7. Only a small fraction of fast and wide CMEs (18%) and DH type II bursts (27%) is associated with SGRE events in SC 24.
8. Hard X-ray bursts at energies > 100 keV have a long duration ($\gtrsim$ 5 min) when accompanying fast and wide CMEs, DH type II bursts, and large SEP events.
9. Hard X-ray bursts at energies > 100 keV when accompanied by DH type II bursts are good indicators of SGRE events evidenced by the observations in SCs 24 and 25.

10. Based on the association with > 100 keV HXR bursts, we estimate that 27 of the 79 DH type II bursts that occurred during the Fermi/LAT data gaps are likely to be SGRE events, bringing the total number of SGRE events in SC 25 to 43.
11. The number of SC 25 SGRE events estimated from the SC 24 association rate between FW CMEs and DH type II bursts is ≈ 48 and 49, respectively, not too different from the estimate based on > 100 keV HXR bursts.
12. If the number of SGRE events followed SSN, we should have ≈ 38 SGRE events in SC 25. However, if the overabundance of FW CMEs and DH type II bursts relative to SSN applies to SGRE, we get 48 and 50 SGRE events, respectively in SC 25.
13. The estimated number of SGRE events in SC 25 is consistent with the observation that SC 25 is stronger than SC 24.

**Funding**

NG is supported by Goddard Space Flight Center, LWS-HSSO, STEREO Project, and NASA Science Mission Directorate, 2024 HISFM. SY, PM, HX, SA are partially supported by Goddard Space Flight Center, PHaSER-670.089. SA and HX are also supported by US National Science Foundation, AGS-2228967.

**Acknowledgments**

The Fermi mission is a joint venture of NASA, the United States Department of Energy, and government agencies in France, Germany, Italy, Japan, and Sweden. We acknowledge the use of the Fermi Solar Flare Observations facility funded by the Fermi GI program (https://hesperia.gsfc.nasa.gov/fermi_solar/). Wind-Konus flare data are from https://www.ioffe.ru/LEA/sun.html. The Global Geospace Science (GGS) Wind is a NASA science spacecraft designed to study radio waves and plasma that occur in the solar wind and in the Earth's magnetosphere. SDO is part of NASA's LWS program. SOHO is a project of international collaboration between ESA and NASA. STEREO is the third mission in NASA's Solar Terrestrial Probes (STP) program. Geostationary Operational Environmental Satellites (GOES) is a collaborative NOAA and NASA program providing continuous imagery and data. The e-CALLISTO data are made available online by the Institute for Data Science FHNW Brugg/Windisch, Switzerland; e-CALLISTO is one of the instrument networks of the international space weather initiative (ISWI). The sunspot number V2.0 is from WDC-SILSO, Royal Observatory of Belgium, Brussels, DOI: https://doi.org/10.24414/qnza-ac80. We thank Elizabeth Hays, Fermi Project Scientist, for discussion on the 2018 solar array drive anomaly.